%% file: S.tex
\documentclass[prb,twocolumn]{revtex4}

\usepackage{color}

\usepackage{latexsym}
\usepackage{amsmath}
\usepackage[latin1]{inputenc}
\usepackage{eucal}
\usepackage{epsfig}
\begin{document}
\newcommand{\dd}{\ensuremath{\text{d}\!\!\:}}
\newcommand{\e}{\ensuremath{\text{e}}}
\newcommand{\HH}{{\ensuremath{\cal H}}}
\newcommand{\U}{{\ensuremath{\cal U}}}
\newcommand{\T}{\ensuremath{\hat T}}
\newcommand{\J}{\ensuremath{\hat J}}
\newcommand{\ket}[1]{\ensuremath{\vert{#1}\rangle}}
\newcommand{\bra}[1]{\ensuremath{\langle{#1}\vert}}
\newcommand{\braket}[1]{\ensuremath{\langle\vert{#1}\vert\rangle}}
\newcommand{\bracket}[2]{\ensuremath{\langle{#1}\vert{#2}\rangle}}
\newcommand{\ds}{\displaystyle}
\newcommand{\transp}{\top}
\title{Nonequilibrium electron transport using the density matrix renormalization group}
\author{Peter Schmitteckert}
\affiliation{Institut f{\"u}r Theorie der Kondensierten Materie,
Universit{\"a}t Karlsruhe, 76128 Karlsruhe, Germany}
\date{\today}
\begin{abstract}
We extended the Density Matrix Renormalization Group method
to study the real time dynamics of interacting one dimensional
spinless Fermi systems by applying the full time evolution operator to
an initial state. As an example we describe the propagation
of a density excitation in an interacting  clean system
and the transport through an interacting nano structure.
\end{abstract}
\maketitle
\section{Introduction}
The density matrix renormalization group method (DMRG) \cite{DMRG:Base,DMRG:Book}
is a powerful technique to study the properties of one dimensional interacting
quantum systems.
The advantage of the DMRG is that it can treat quantum lattice systems
in the presence of site-dependent interaction, hopping parameter, and
on-site potentials \cite{DMRG:Disorder} with high accuracy, including
subtle lattice effects like multiple umklapp processes.\cite{DMRG:MUS}
\par
Originally the  method was set up to describe the
equilibrium properties of the ground state and a few excited states.
It was then extended to calculate frequency dependent spectral functions
by the use of the $[ H - E_0 - \omega + \imath\eta]^{-1}$ operator.%
\cite{DMRG:SpktrlFkt-Hallberg,DMRG:SpktrlFkt-White,%
DMRG:SpktrlFkt-Jeckelmann}
\par
A second approach to study transport properties within the framework
of DMRG is to relate equilibrium properties of the ground state to
transport properties. Molina {\em et al.}\ \cite{DMRG:PC-Molina}
and Meden and Schollwöck \cite{DMRG:PC-Meden}
calculated the conductance through an interacting nano structure attached
to leads by relating the conductance of the system to the ground state curvature,
based on an idea by Sushkov.\cite{Sushkov01}
\par
Cazalilla and Marston \cite{TdDMRG:Marston} used the basis states of
the last DMRG sweeps to integrate the Schrödinger equation in real time.
However,  as shown by Luo, Xiang, and Wang,\cite{TdDMRG:MarstonComment}
their approach was inconsistent with the DMRG scheme, since they used a basis
to integrate the Schrödinger equation which was only adapted to the $t=0$ state,
neglecting the relevant states for $t>0$ during the DMRG sweeps.
\section{Calculation of the time evolution}
Instead of integrating the time dependent Schrödinger equation numerically,
we make use of the formal solution and apply the full time evolution operator
\begin{equation} \label{eq:ExpU}
	\U(t_2,t_1) = \e^{ -\imath \HH ( t_2 - t_1) }
\end{equation}
to calculate the time dependence of an initial state $\ket{\xi(0)}$
\begin{equation}
	 \ket{\xi(t)} = \U(t,0) \ket{\xi(0)} \,,
\end{equation}
where \HH\ is the Hamiltonian of the system of interest.
\par
While the calculation of $\e^{ -\imath \HH t}$ is not feasible for large
matrix dimensions, one can calculate the action of a matrix
exponential on a vector similar to the diagonalization of sparse matrices,
where one cannot diagonalize the full matrix,
but one can search for selected eigenvalues and eigenvectors.
We apply a Krylov subspace approximation \cite{Saad98}
to calculate the action of the matrix exponential in
equation (\ref{eq:ExpU}) on a state $\ket{\xi}$.
We would like to encourage the reader interested in implementing a matrix
exponential to study the excellent review by Moler and Van Loan \cite{MolerLoan03}
and discourage the use of a Taylor expansion.
In our implementation we make use of a Pad{\'e} approximation
from {\sc Expokit} \cite{Expokit} to calculate the dense matrix exponential
in the Krylov subspace.
In order to calculate $\ket{\xi(t)}$ up to a final  time $T$,
we discretize the time interval into $N$ time steps
$\{ t_0, t_1, t_2, \cdots, t_N \}$ with $t_0=0$, $t_N = T$  and $t_j < t_{j+1}$,
typically $ t_j - t_{j-1} = 0.5$ and $N = 2T$.
It turns out that using a time slice $t_j - t_{j-1}$ of the order of one
is sufficient to ensure a fast convergence of
\begin{equation}
  \ket{\xi(t_j)} = \e^{ -\imath \HH ( t_j - t_{j-1}) } \ket{\xi(t_{j-1})}
 \end{equation}
It is crucial that one does not need to store the wave function at all time steps.
Instead one can add $\ket{\xi(t_j)}$ immediately to the density matrix $\rho$
and calculate the matrix elements of interest at each time step separately.
Therefore, only the two wave functions
$\ket{\xi(t_j)}$ and $\ket{\xi(t_{j-1})}$
are needed during the calculation of the time evolution.\cite{Note1}
\par
One even does not need to calculate the ground state of the unperturbed system,
it is sufficient to keep the matrix elements of the \HH\, which may be useful
for systems with highly degenerate ground states, e.g., an XXZ ferromagnet,
where the convergence of the diagonalization is very slow.
However, in this work we calculate the $m$ lowest lying eigenstates $\ket{\Psi_m}$
of $\HH$, $E_m \ket{\Psi_m} = \HH   \ket{\Psi_m}$,
and use the time evolution operator only on the subspace orthogonal
to these eigenstates,
\begin{eqnarray}
	\hat{P} &=& \sum_{m=0}^{m-1} \ket{\Psi_m}\bra{\Psi_m} \label{eq:Projector}\\
	\ket{\xi(t)} &=& \sum_{m=0}^{m-1}  \e^{-\imath (E_m-E_0) t} \ket{\Psi_m} \bracket{\Psi_m}{\xi(0)} \\ \nonumber
	 & & \;+\; \e^{-\imath (\HH-E_0) t }  ( 1 - \hat{P}) \ket{\xi(0)} \,,
\end{eqnarray}
and calculate the dynamics in the subspace
of the eigenstates $\{ \Psi_0, \ldots, \Psi_{m-1} \}$ exactly.
In addition we introduced  a phase  choice $ \e^{\imath E_0 t }  $ to make the ground state time independent.
\begin{figure*}
\begin{center}
                        \epsfig{file=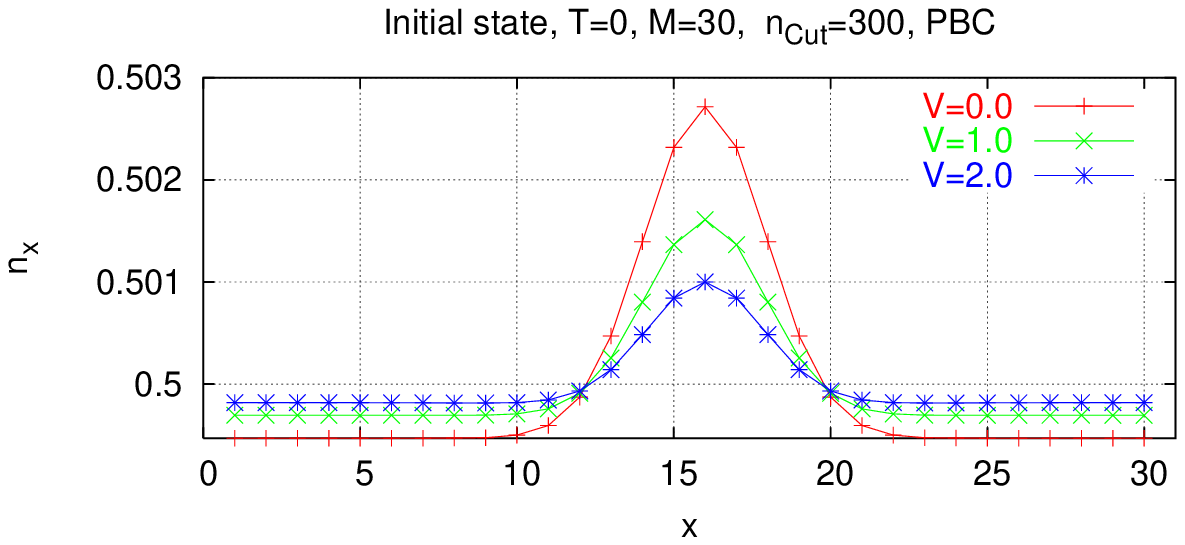, scale=0.65} \hspace{3ex}
                        \epsfig{file=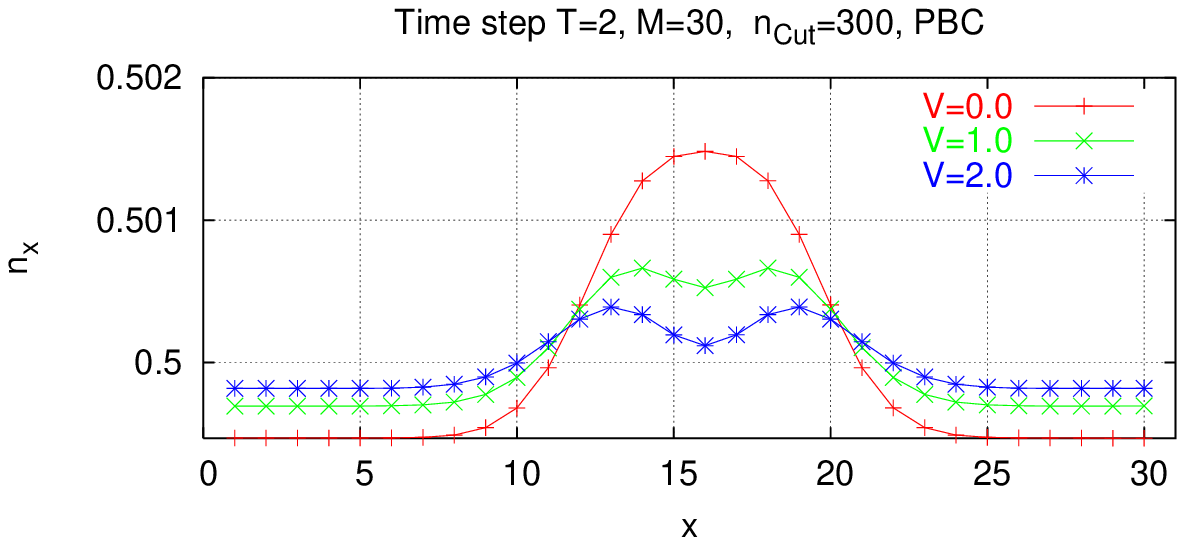, scale=0.65}\\
                        \epsfig{file=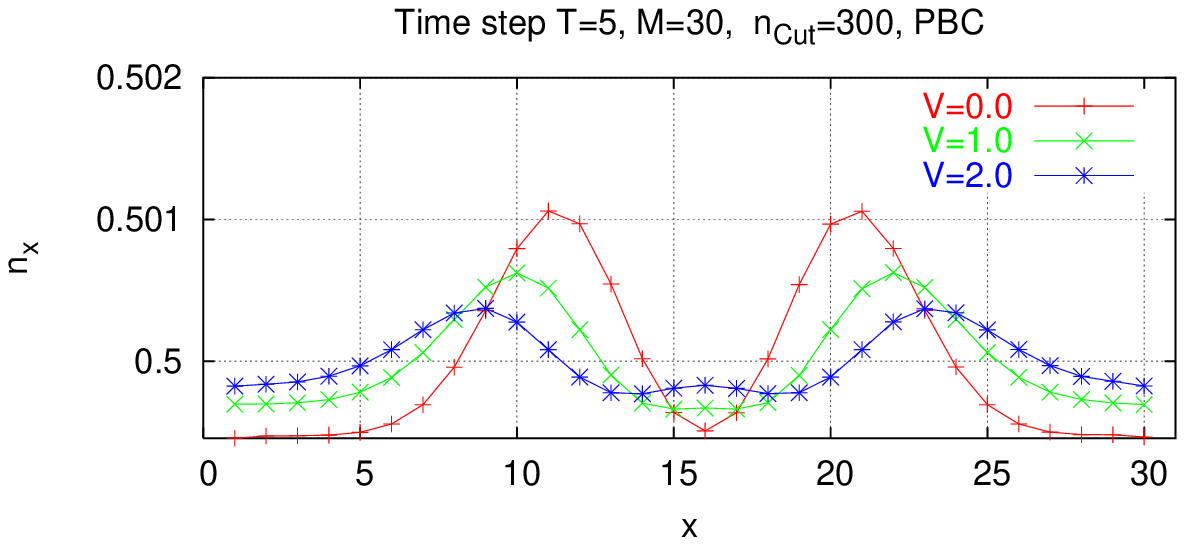, scale=0.65} \hspace{3ex}
                        \epsfig{file=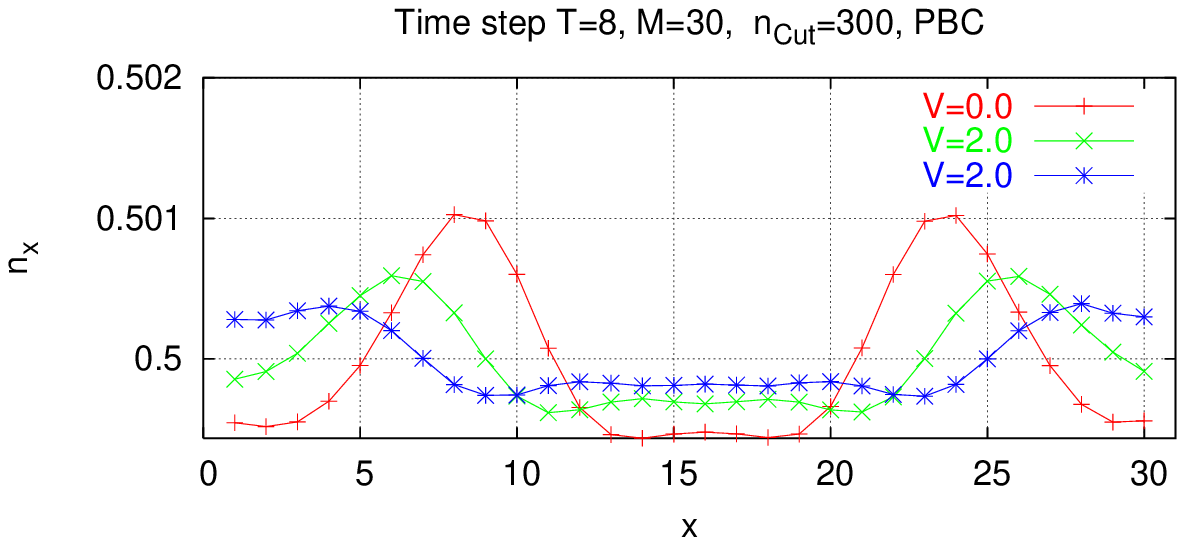, scale=0.65}
\end{center}
\caption{ Time evolution of a wave packet for a system of $M=30$ sites,
periodic boundary conditions, and $V=0.0$ (plus), $V=1.0$ (crosses) and
$V=2.0$ (stars).
The snapshots are taken at $T=0$, $2$, $5$, and $8$.
$n_{\rm Cut}=300$ states per block were kept within the DMRG procedure.}
\label{fig:Waves_Clean_PBC}
\end{figure*}
\section{Wave packet dynamics}
In order to prepare an initial state we apply a small perturbation $\delta\HH$ to
the Hamiltonian \HH\ of the system of interest  and calculate
$\ket{\xi(0)}$ as the ground state of $\HH + \delta\HH$.
\par
As a first example we study the time evolution of a density pulse in a model
of interacting spinless fermions:
\begin{equation} \label{eq:SpinlessFermionsGauss}
\HH = -t\,\sum_{x=1}^{M} c^\dagger_{x} c^{\phantom{\dagger}}_{x-1} +c^\dagger_{x-1} c^{\phantom{\dagger}}_{x}
    \;+\; V\,\sum_{x=1}^{M}  \,n_{x} n_{x-1}
\end{equation}
where $t$ is the hopping parameter and $V$ the nearest neighbour interaction parameter.
In this work we measure all energies with respect to $t=1$.
To create a wave packet we add a Gaussian potential
\begin{equation} \label{eq:SpinlessFermions}
 \delta\HH =   \;-\; \mu \sum_{x=1}^{M}  \e^{  -\frac{ (x-x_1)^2}{2\sigma^2}} \, n_{x}
\end{equation}
where $\mu$ is the strength, $\sigma$ the width, and $x_1$ the position
of the perturbation.
\begin{figure}
\begin{center}
                        \epsfig{file=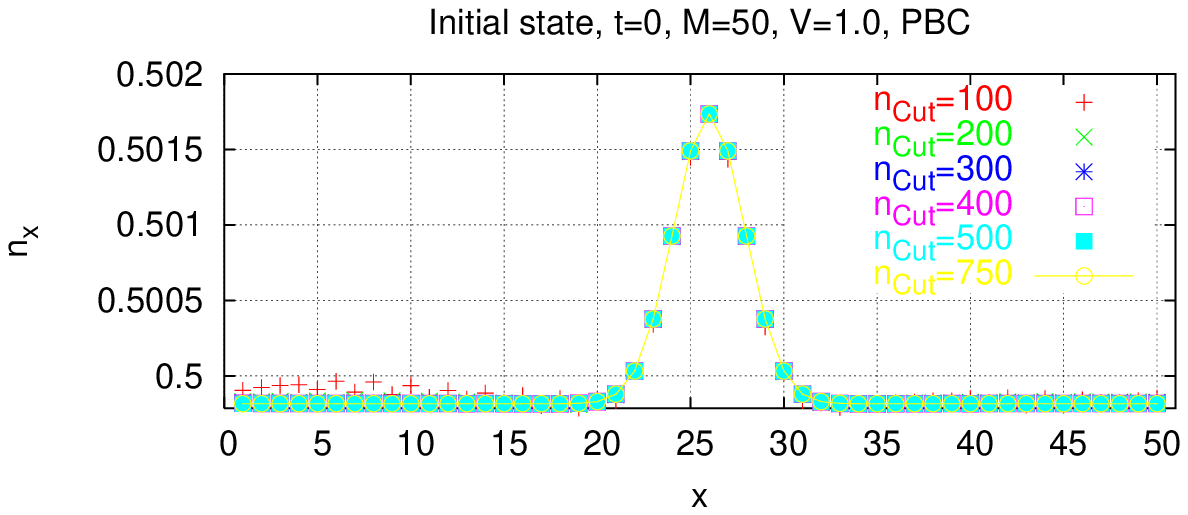, scale=0.7} \\
                        \epsfig{file=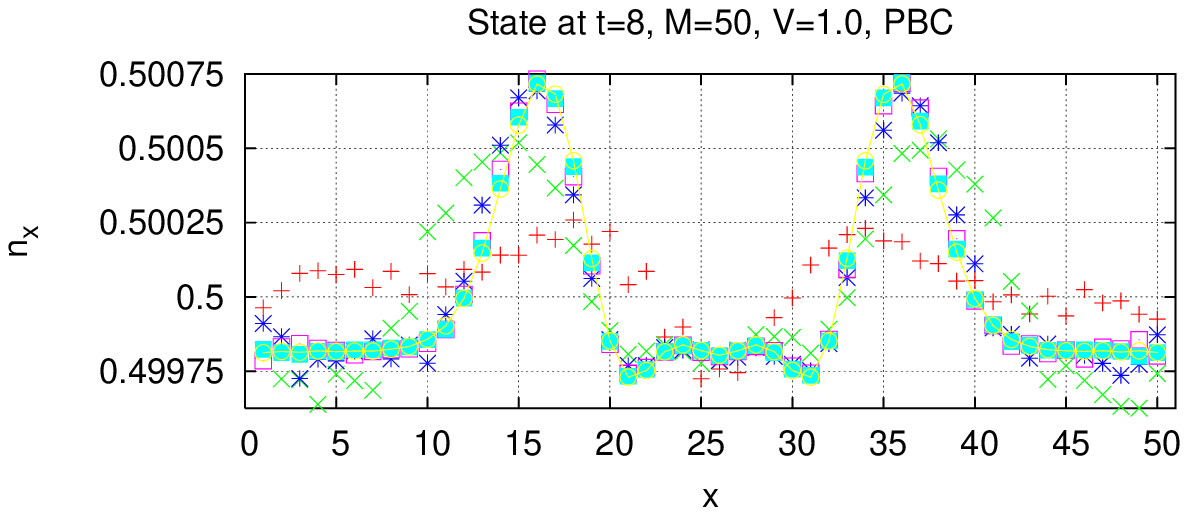, scale=0.7}
\end{center}
\caption{ (a) Initial wave packet for a 50 site system with periodic boundary
conditions and $V=1$, for different numbers of states kept per block: $n_{\rm Cut}=100$ (plus),
$200$ (crosses), $300$ (stars),  $400$ (open squares), $500$ (filled squares)
and $750$ (open circles).
(b) Same system at time $T=8$.}
\label{fig:Waves50_m}
\end{figure}
In Fig.\ (\ref{fig:Waves_Clean_PBC}) we have plotted the time evolution of an
initial wave packet at $x_1=6$ in a 30 site system at half filling
and periodic boundary condition using 300 states per DMRG block.
Due to time reversal symmetry the initial state consist of a left and a right moving
wave packet. During the time evolution the initial peak splits into two peaks which
are moving with the group velocities $\pm v_g$.
For $V=0$ the DMRG results coincides with the result
from exact diagonalization.
As a first result this method gives direct access to the
group velocity $v_g$ of  a density excitation
without relying on finite size analysis or arguments from conformal field theory.\cite{BA:QISM}
In table (\ref{tbl:vF}) we compare the extracted group velocities $v_g$
for the $M=30$ site system with the Fermi velocity $v_F$ known from Bethe ansatz results
for an infinitesimal excitation in the infinite system size limit,
$ v_F = \pi \sin(2\eta)/ (\pi - 2\eta)$ with the usual parametrization
$V = -2\cos(2\eta)$.\cite{BA:QISM}
As expected from $v_F$ the wave packets travel
faster the stronger repulsive interaction are,
while they are slowed down by attractive interaction.
For $\mu=0.002$ there is a good agreement from $v_g$ with $v_F$,
while the results for $\mu=0.02$ already include dispersion effects.
In addition, the broadening of the wave packets reveals information
on the dispersion relation. A detailed study is beyond the scope of this work and subject
for future studies.
\begin{table}
\begin{tabular}{|c|c|c|c|c|c|}\hline\hline
	$U$ 			&   -1.5 		& 0.0 	& 0.5 & 1.0 & 1.5 \\\hline
	$v_g$, $\mu=0.02$&    1.0		& 1.9 	& 	2.2&	 2.47 &2.71 \\\hline
	$v_g$, $\mu=0.002$&   0.92		& 2.00	& 	2.30&	 2.59 &2.87  \\\hline
	$v_F$(BA) 	&  0.88  	& 2.00 	& 	2.31&	2.60&2.88  \\ \hline\hline
\end{tabular}
\caption{ Comparison of $v_g$ extracted from DMRG simulations for a $M=30$ site system
and a potential strength $\mu = 0.02$ , $\mu=0.002$ and Bethe ansatz results for $v_F$ in the infinite system
and an infinitesimal small excitation.}
\label{tbl:vF}
\end{table}
\par
It is not obvious that one can target for a few low lying states
$\ket{\Psi_m}$ of \HH,
the ground state $\xi(0)$ of $\HH + \delta\HH$ and $N\sim100$ time steps of  $\xi(t_j)$
simultaneously in each DMRG step.
However, since the DMRG truncation  is the only approximation
in our method, we can systematically increase the number of states $n_{\rm nCut}$ kept per
block to control errors due to the Hilbert space truncation.
In Fig.~(\ref{fig:Waves50_m}) we plot an initial wave packet  $\xi(0)$ and
the wave packet $\xi(T)$ at $T=8$ for a 50 site system, a potential strength of $\mu=0.02$,
 an interaction strength $V=1.0$ and periodic boundary conditions.
 While for the initial state 200 states per block seem to be sufficient to describe the wave packet,
 far more states are needed to obtain the dynamics of the wave packet correctly.
 The slow convergence at the boundaries of the system is related to an implementation detail.
 We do not keep all density operators to evaluate $n_x$ in the final iteration step.
 Instead we calculate $n_x$ during the last 1.5 finite lattice sweeps, when we have
 the operators available. The advantage of this procedure is that one does not need to keep
 all operators at the price that the operators close to boundaries have less accuracy, since they are
 evaluated in a highly asymmetric block configuration.
 \par
 Remarkably, the overlap $\bracket{\Psi_0}{\xi(0)}$ between the ground state of \HH\ and
 the ground state  of the system of $\HH + \delta\HH$ shown in Fig.~(\ref{fig:Waves50_m})
 is $99.99\%$. Therefore, it is the
 $0.01\%$ contribution which gives the initial excitation and governs the time evolution.
 This high overlap was the motivation to introduce the projection defined im Eq.~(\ref{eq:Projector}).
\section{Transport through a quantum dot}
In order to study transport through an interacting nano structure,
we prepare a system consisting of an interacting region coupled to
noninteracting leads, see Fig.~(\ref{fig:NanoSystem}),
\begin{eqnarray} \label{eq:SpinlessFermionsNanoSystem}
\HH &=& -t\,\sum_{x=1}^{M} \left( c^\dagger_{x} c^{\phantom{\dagger}}_{x-1}
	+c^\dagger_{x-1} c^{\phantom{\dagger}}_{x} \right)
    \;+\; U\,\sum_{x=n_1+1}^{n_2-1}  \,n_{x} n_{x-1} \nonumber\\
    & & \;+\; \gamma\,U\,\sum_{x=n_1, n_2}  \,n_{x} n_{x-1}  \,,
\end{eqnarray}
where $t$ is hopping parameter, $U$ is the interaction on the nano structure and
$\gamma$ defines a smoothening of the onset of interaction at the nano structure.
In this work we have set $\gamma=0.5$, compare Molina {\em et al.}\ \cite{DMRG:PC-Molina}.
In the following we denote $M_S= n_2 - n_1$ the number of sites in the nano structure,
$M$ the number of site of the total system and $M_L = M - M_S$ the number of lead sites.
\par
In Fig.~(\ref{fig:Waves_Dot_OBC}) we show the time evolution of a wave packet initially
placed in the left lead of a system with $M_S=7$,  $M=50$, interaction $V=0.0$, $1.0$, $2.0$
and $5.0$ and hard wall boundary conditions, which lead to
perfect reflection at the chain ends. To rule out truncation errors we use $n_{\rm Cut} =1000$
states per block, compare discussion of Fig.~(\ref{fig:Waves50_m}).
We have averaged the density over neighbouring sites to smoothen out the
Friedel oscillations (for all $V$) and the charge density wave on the dot for $V=5.0$.
At the beginning of the time evolution, the wave packet is not overlapping
with the interacting  nano structure, hence
the packets travel synchronously for all interaction strength.
Once they reach the nano structure, the wave packets move with the group velocity
of the interacting system. After the wave packets have left the interacting region they continue
to move at the velocity of the noninteracting system. The wave packets
which traveled through the interacting region are now travelling in front of those
with smaller interaction.
\par
For $U<=2.0$ the nano structure is transparent, although there seems to be a reflection
of a negative pulse as predicted by Safi and Schulz \cite{Safi95}.
For  very strong interaction, $U>2.0$, there is an instability to a
charge density ordering,\cite{DMRG:MUS,BA:QISM} and the nano structure has a finite reflection.
We would like to remark that these simulations clearly demonstrate that the Luttinger description
of the infinite system makes already sense for a system consisting of a few lattice sites only.
One should should keep in mind that for such small systems the effective parameters%
, like $v_F$,
have not reached the infinite system limit.
However, the scaling already leaves its fingerprint.
 \begin{figure}
\begin{center}
\input{NanoSystem.tex}
\end{center}
\caption{Nano structure attached to leads.}
\label{fig:NanoSystem}
\end{figure}
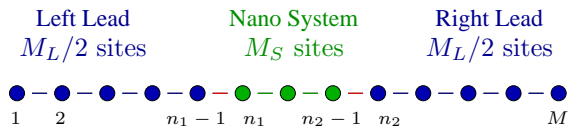
\begin{figure*}
\begin{center}
                        \epsfig{file=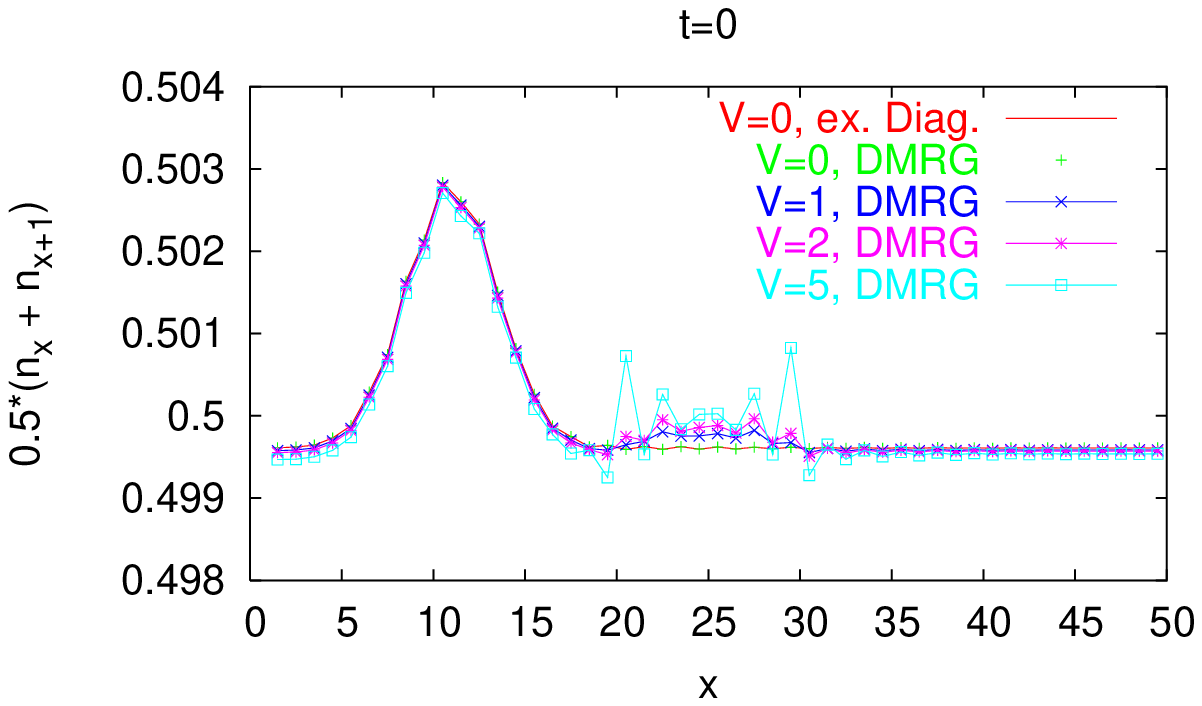, scale=0.65} \hspace{3ex}
                        \epsfig{file=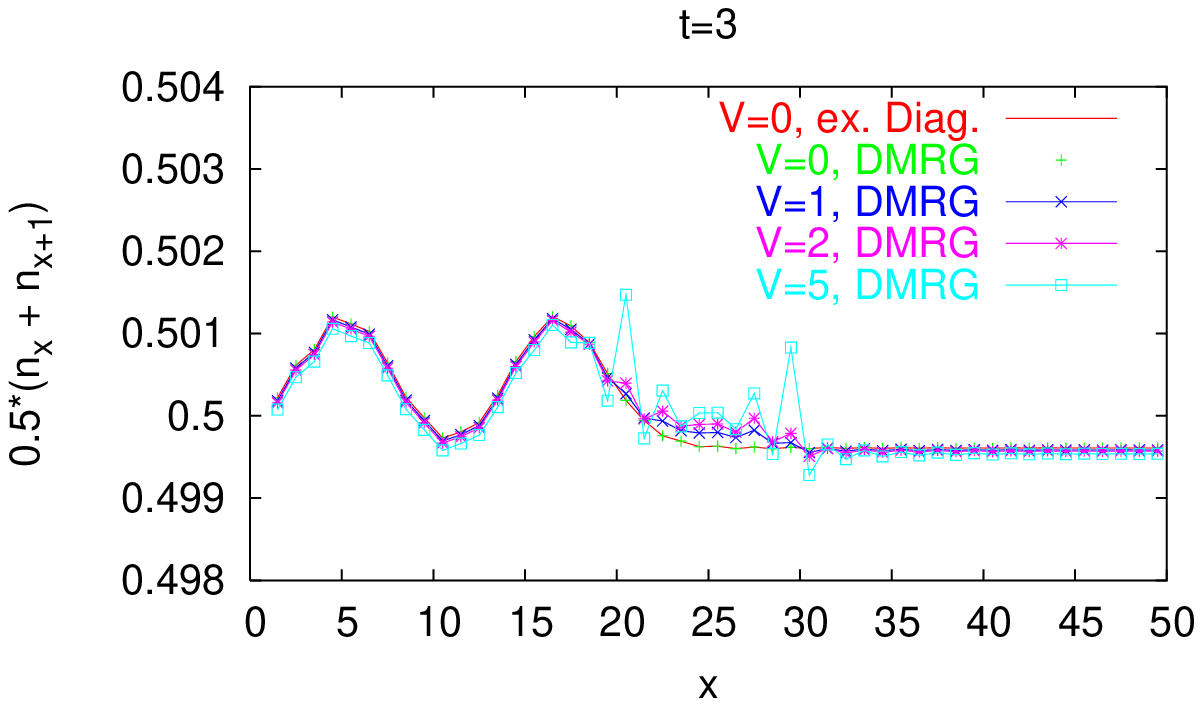, scale=0.65}\\
                        \epsfig{file=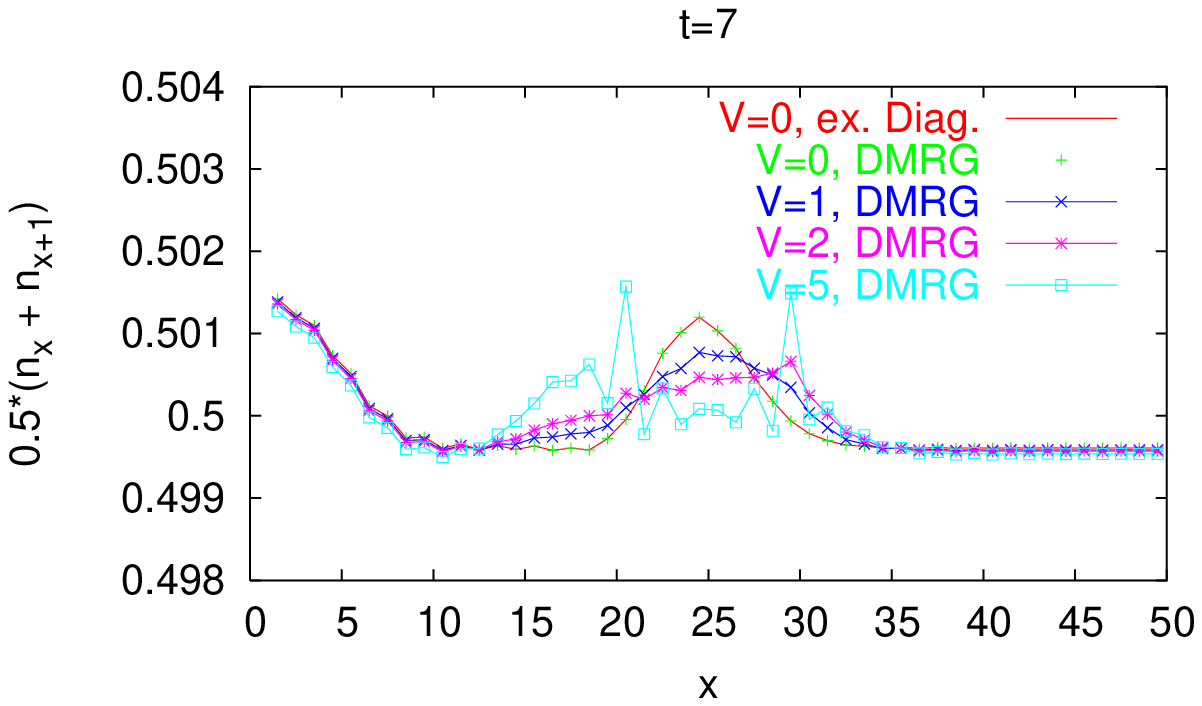, scale=0.65} \hspace{3ex}
                        \epsfig{file=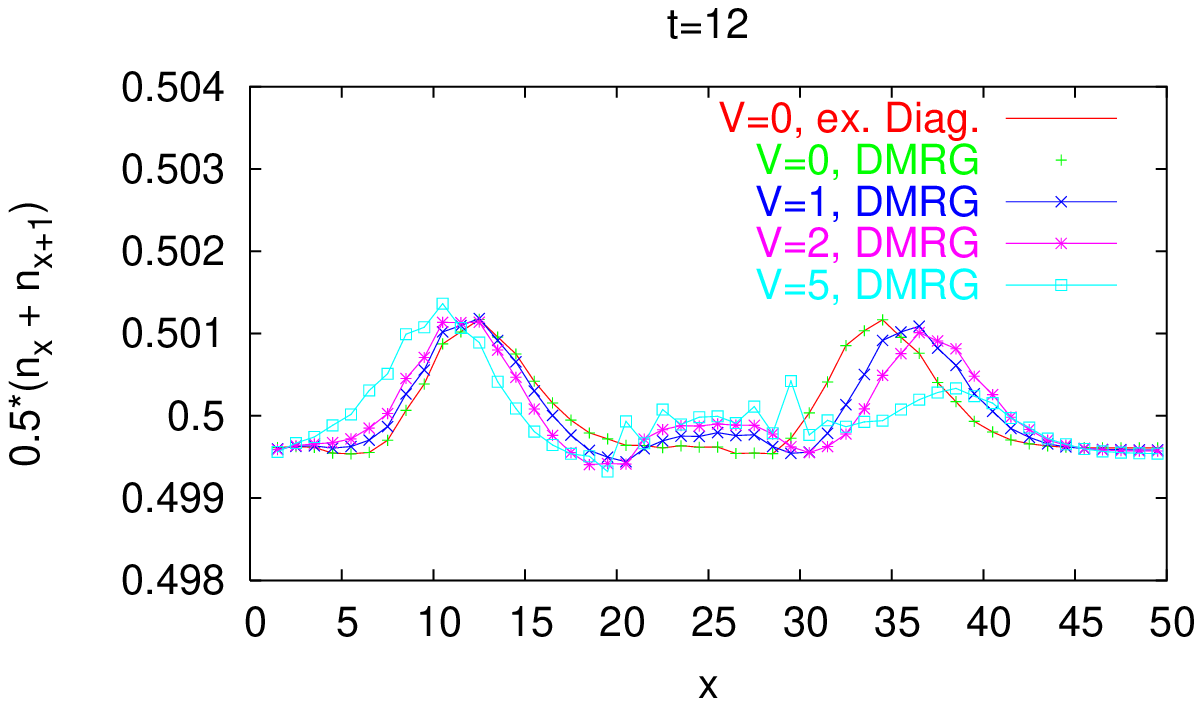, scale=0.65} \\
\end{center}
\caption{ Transport through an interacting region of $M_S=7$ sites and  $M_L=43$ lead sites,
hard wall boundary conditions and $n_{\rm Cut}=1000$.
The snapshots are taken at $T=0$, $3$, $7$, $12$.
The density is averaged over 2 neighbouring site to smoothen $2k_F$ oscillations
and is plotted vs.\ the site location $x$ for $V=0$ (plus), $V=1$ (crosses), $V=2$ stars,
and $V=5$ (squares). For $V=0$ the line is calculated by an exact diagonalization.}
\label{fig:Waves_Dot_OBC}
\end{figure*}
\section{Summary}
In summary we have shown an accurate method to calculate real time dynamics
within the framework of DMRG. By applying the matrix  exponential on our initial state
state we are able to perform the complete time integration
of the time dependent Schrödinger equation in each single DMRG step.
In this setup the only approximation is given by the truncation procedure
of the DMRG, which be can systematically checked by increasing the number of states
kept during the DMRG sweeps.
\section{Note}
While preparing this work we became aware of related work \cite{TdDMRG:White,TdDMRG:Schollwoeck}
on using real time dynamics within the DMRG.
Both apply a Suzuki-Trotter decomposition of the time evolution operator,
based on an work by  by Vidal.\cite{Vidal03}
In addition, their work relies on the state prediction \cite{DMRG:WhitePrediction}
to calculate the time evolution of a state, which represents an additional
approximation.
In contrast to their work we calculate the initial state
and apply the full time evolution operator in each iteration step,
without introducing additional approximations beyond the truncation scheme of the DMRG.\\[1pt]

\section{Acknowledgement}
I acknowledge the support of the Center for Functional Nano Structures
within project B2.10. I would like to thank Karl Meerbergen for his hints on the
matrix exponential and insightful discussions with Ralph Werner, Peter Wölfle
and Gert L.~Ingold.

\end{document}

%% file: NanoSystem.tex
\begin{picture}(0,0)%
\includegraphics{NanoSystem.pstex}%
\end{picture}%
\setlength{\unitlength}{829sp}%
\begingroup\makeatletter\ifx\SetFigFont\undefined%
\gdef\SetFigFont#1#2#3#4#5{%
  \reset@font\fontsize{#1}{#2pt}%
  \fontfamily{#3}\fontseries{#4}\fontshape{#5}%
  \selectfont}%
\fi\endgroup%
\begin{picture}(17108,3601)(1633,-3134)
\put(11476,-2986){\makebox(0,0)[b]{\smash{{\SetFigFont{7}{8.4}{\rmdefault}{\mddefault}{\updefault}{\color[rgb]{0,0,0}$n_2-1$}%
}}}}
\put(7426,-2986){\makebox(0,0)[b]{\smash{{\SetFigFont{7}{8.4}{\rmdefault}{\mddefault}{\updefault}{\color[rgb]{0,0,0}$n_1 - 1$}%
}}}}
\put(9181,-2986){\makebox(0,0)[b]{\smash{{\SetFigFont{7}{8.4}{\rmdefault}{\mddefault}{\updefault}{\color[rgb]{0,0,0}$n_1$}%
}}}}
\put(13231,-2986){\makebox(0,0)[b]{\smash{{\SetFigFont{7}{8.4}{\rmdefault}{\mddefault}{\updefault}{\color[rgb]{0,0,0}$n_2$}%
}}}}
\put(10351,-961){\makebox(0,0)[b]{\smash{{\SetFigFont{10}{12.0}{\rmdefault}{\mddefault}{\updefault}{\color[rgb]{0,.56,0}$M_S$ sites}%
}}}}
\put(18226,-2986){\makebox(0,0)[b]{\smash{{\SetFigFont{7}{8.4}{\rmdefault}{\mddefault}{\updefault}{\color[rgb]{0,0,0}$M$}%
}}}}
\put(16201,-961){\makebox(0,0)[b]{\smash{{\SetFigFont{10}{12.0}{\rmdefault}{\mddefault}{\updefault}{\color[rgb]{0,0,.56}$M_L/2$ sites}%
}}}}
\put(4051,-961){\makebox(0,0)[b]{\smash{{\SetFigFont{10}{12.0}{\rmdefault}{\mddefault}{\updefault}{\color[rgb]{0,0,.56}$M_L/2$ sites}%
}}}}
\put(3376,-2986){\makebox(0,0)[b]{\smash{{\SetFigFont{7}{8.4}{\rmdefault}{\mddefault}{\updefault}{\color[rgb]{0,0,0}$2$}%
}}}}
\put(2026,-2986){\makebox(0,0)[b]{\smash{{\SetFigFont{7}{8.4}{\rmdefault}{\mddefault}{\updefault}{\color[rgb]{0,0,0}$1$}%
}}}}
\end{picture}%